# Measurement of Magnetic Relaxation in the peak regime of $V_3Si$

G. Ravikumar[1], M. R. Singh[1] and H. Küpfer[2]

[1]Technical Physics and Prototype Engineering Division
Bhabha Atomic Research Centre, Mumbai – 400085, India

[2]Institut für Festkörperphysik, Forschungszentrum Karlsruhe and Universität Karlsruhe, 76021 Karlsruhe, Germany

Abstract

Magnetization relaxation measurements are carried out in the Peak effect regime of superconducting $V_3Si$ crystal, using Quantum Design SQUID magnetometer. Relaxation in the increasing field scan is logarithmic in time, consistent with the theory of flux creep. The relaxation on the decreasing field scan however exhibits athermal behavior which is predominantly governed by the flux avalanches triggered by the small external field perturbation experienced by the superconductor during measurement scan in an inhomogeneous field.

## 1. Introduction.

Magnetization relaxation measurements have been the cornerstone of vortex dynamics studies in the mixed state of type II superconductors [1]. They were extensively used near the second magnetization peak (SMP) in high $T_c$ superconductors such as $YBa_2Cu_3O_{7-x}$[2,3] and $Bi_2Sr_2CaCu_2O_8$[4,5]. SMP in these materials represents a vortex matter transition from a quasi-ordered Bragg Glass to highly disordered Vortex Glass phase. In low $T_c$ materials, this transition is marked by a sharp peak in critical current density $J_c$ as a function of field and temperature (peak effect). However, to the best of our knowledge, there has not been any relaxation study in the peak effect regime of low $T_c$ superconductors, despite widespread interest in the phenomenon[6]. Recently Pei-Chun Ho *et al*[7] reported magnetization relaxation data in the peak effect regime of $CeRu_2$ using Quantum Design (QD) SQUID magnetometer. They observed quite a rapid decay of induced currents. At higher temperatures the current decay is almost complete within few hours. Further, relaxation rate on the decreasing field branch of the magnetization hysteresis loop is found to be much larger than that on the increasing field branch. In other words, relaxation rate is asymmetric with respect to the direction of field scan.

Below the field $H_p$ (where $J_c$ is maximum), pronounced history dependence is observed in $J_c$ [8-13]. Each value of $J_c$ produced under different field - temperature history, corresponds to a macroscopically distinguishable metastable configuration of the vortex lattice[14]. Lower $J_c$ corresponds to a configuration where vortex lattice correlations extend over larger distances[15]. Magnetization relaxation in the peak effect regime may be governed by two different processes. Thermally activated flux creep, where individual vortices or vortex bundles hop over pinning barriers, always results in the decay of magnetization currents[16]. Secondly, a metastable configuration may thermally overcome the free energy barrier in the configuration space to relax to a lower energy configuration, which, of course, need not correspond to a lower $J_c$[14,17,18]. It is argued that such a relaxation process is highly unlikely in low $T_c$ materials, because the barriers separating different metastable configurations are too large compared to thermal energy $kT$[14]. However, any perturbation causing vortex motion can drive the vortices in a metastable configuration into a stable state[17,18].

In this paper, we present magnetic relaxation measurements in the peak effect regime of $V_3Si$ using QD SQUID magnetometer. The most important result is, the relaxation of the supercooled vortex phase on the decreasing field branch[11,12], is not governed by any of the thermally activated relaxation processes discussed above. It is predominantly due to the field excursion experienced by the sample during the measurement process[19].

## 2. Experiment.

The $V_3Si$ crystal (1.6mm × 0.5mm × 0.3mm) used in this study has a superconducting transition temperature $T_c$ = 16.5K. Magnetic measurements are carried out using a QD SQUID magnetometer with a maximum field of 5.5T. The field is oriented along the smallest dimension of the sample. The sample exhibits a prominent peak effect below 4 T at a temperature 14.5K. Use of two different scan lengths, viz., 2 cm and 1.6 cm, produce almost identical hysteresis loops as shown in Fig. 1. The field inhomogeneity for the two scan lengths differs by a factor of approximately 2.4 [19]. All the measurements are performed in the fixed range mode and the sample is scanned only once for each data point. The asymmetry observed in the magnetization on the increasing (forward) and decreasing (reverse) field scans is due to the history dependent $J_c$ [12].

Relaxation measurements are carried out at different positive fields on the forward and reverse field branches of the hysteresis loop. In the forward case, the initial magnetic moment is recorded after increasing the field from -1.0 T to the target field to erase the previous history. In the reverse case the field is brought down from 5 T (larger than the second critical field) to the target field. At each field, magnetic moment is measured as a function of time using two to five different values of regular time intervals to check whether the process of measurement itself influences the magnetic moment. More extensive data is taken on the reverse curve. Relaxation between 2 – 2.5 T on the forward curve is not reliable because magnetic moment is too small to give good regression.

## 3. Results

*(a) Forward relaxation.*

Fig. 2 shows two sets of magnetic moment $m$ *vs* time $t$ data recorded at 2.7 T on the forward branch of the hysteresis loop at regular time intervals of 100s and 300s. Actually the time interval between two measurements is larger by 35 seconds after accounting for the duration of the measurement scan. In each case, 40 measurements were recorded with a total duration of about 1.5 and 3.5 hours respectively. Both the data sets fall on the same curve, quite in contrast to the behavior on reverse magnetization curve as would be shown below. A good fit to the logarithmic behavior indicates that the phenomenon of thermally activated flux creep is responsible for the relaxation. Inset of Fig. 2, depicts the field dependence of the normalized logarithmic relaxation rate $S = (1/m_0)\ (dm/dlnt)$ across the peak effect, where $m_0$ is the magnetic moment recorded in the first measurement. $S$ is minimum at the field $H_p$ where hysteresis ( or $J_c$) is maximum.

*(b) Reverse relaxation.*

In the main panel of Fig. 3(a) we show two sets of *m vs t* data recorded at regular intervals of 200 and 600 seconds at a field of 2.7T on the reverse branch. Surprisingly, when the magnetic moment *m* is plotted as a function of the serial number of the measurement *n* (rather than *t*), the two data sets collapse on to the same curve, in spite of the total duration of the relaxation being widely different in the two cases (about 2.6 and 7 hours respectively). For comparison, inset of Fig. 3(a) shows the same data plotted in the form of *ln(m) vs ln(t)* curves. The data in the inset of Fig. 3(a) is replotted in Fig. 3(b) with the 200s interval data moved to the right by $ln(t_0) = 0.99$, so that it overlaps with the 600s interval data. These results clearly indicate that the observed relaxation is not governed by any thermally activated process. We call this *athermal* relaxation. Magnetic moment falls sharply by about 20-30% in the first two/three measurements followed by a few percent change in the subsequent measurements. In a purely athermal relaxation, we note that $ln(t_0)$ should equal $ln(635/235) = 0.994$. On the other hand, if the relaxation is purely of thermal origin as in the forward case, $ln(t_0) = 0$ signifying that the two data sets must collapse on to the same curve anyway. We argue that $0 < ln(t_0) < 1$ signifies a combination of thermal and athermal relaxation processes.

At 3.3T (above $H_p$), neither *m vs n* data nor *m vs t* data for the two time intervals overlap exactly. In Fig. 4, we plot the *ln(m) vs ln(t)* by shifting the 200s interval data by $ln(t_0) = 0.62$, indicating that the relaxation is now a combination of thermal and athermal processes. It is important to note that only the long term relaxation data can be made to overlap with each other by such a shift along the time axis. For the shorter time data to overlap, $ln(t_0)$ must again be approximately 0.99. This suggests that the short term relaxation is governed by athermal process while the long term relaxation is only partially governed by thermally activated process. We may therefore argue, even at lower fields, the relaxation process may perhaps cross over into a thermally activated process at very long time scales. $ln(t_0)$ plotted as a function of applied field (Fig. 5) now characterizes the process of relaxation. Below $H_p$, the relaxation is purely athermal, while the thermal component gradually increases above the peak. Surprisingly, athermal relaxation process persists even at 3.4 T.

Fig. 6 compares *m vs n* curves measured at 200s intervals at a field of 2.4 T on the reverse curve using 1.6 and 2.0 cm scan lengths. A sharp drop in *m* occurs in the first few measurements in both the cases. The relaxation recorded with 1.6 cm scan is slightly less than that obtained using 2 cm scan length. In section 4, we shall argue, that the athermal relaxation is due to the measurement process employed in the QD SQUID magnetometer.

## 4. Discussion.

In a SQUID magnetometer, magnetic dipole response across a pickup coil in the second order gradiometer configuration is measured by scanning the sample along its axis. However field inhomogeneity along the scan is described by [19]

$$H(z) \approx H_a\,[1 - 1.25 \times 10^{-4}\, z^4\,],$$

where $z$ is the distance measured from the center of the scan and $H_a$ is the field at $z = 0$. For instance, a 2 cm scan begins at $z = -1$ cm where the sample actually sees a field ($H_a - \delta H_a$ ) rather than $H_a$, before the measurement begins (see Fig. 7(a)). It experiences a field cycle of amplitude $\delta H_a \approx 1.25 \times 10^{-4} H_a$ as it moves up along the axis upto $z = +1$ cm (see Ref. 19). To describe the evolution of critical state flux profile at a given field on the reverse magnetization curve, let us consider a superconducting slab extending from 0 to $2a$ in the $x$-dimension and infinite in the other two directions. Field is assumed to be applied perpendicular to the x- dimension. Symmetry with respect to $x = a$ permits us show only half the slab ($0 < x < a$).

The critical state on the reverse magnetization curve before a measurement commences, is described by the field profile A with boundary condition $H(x = 0) = ( H_a - \delta H_a )$. The field excursion experienced by the sample in a measurement is equivalent to a field cycle of amplitude $\delta H_a$ at the surface of the slab. In the peak effect region of our $V_3Si$ crystal (at 14.5 K), the field perturbation required to penetrate the center of the sample is of the order of few 100 G. But the field perturbation experienced by the sample during a measurement scan is approximately 4 G in a 2 cm scan and less than 2 G in a 1.6 cm scan. Such a perturbation penetrates only a small depth $p$ near the surface (see the modified field profile within the depth p in Fig. 7(a) with the boundary condition $H(x = 0) = H_a$ ). It is expected not to affect the measured magnetization significantly[19]. However, in the peak effect regime, the vortex state on the reverse magnetization curve is a supercooled high $J_c$ phase[12]. The oscillatory field perturbation drives the vortices into a relatively more ordered configuration resulting in a sharp fall in $J_c$ (and thus the slope of the flux profile) in the surface region[17]. This drastically affects the flux profile in the interior of the slab as shown by the dotted line in Fig. 7(a), leading to an avalanche of flux exiting the superconductor. The total flux exiting in an avalanche is equal to the area between the profiles A and B. The modified flux profile B after one field oscillation is now a superposition of two linear regions as depicted in Fig. 7(a), a low $J_c$ region near the surface and a high $J_c$ region in the interior. Actually, the change in the local field slightly modifies the $J_c$ in the interior as well. The oscillatory field perturbation penetrates progressively deeper as $J_c$ in the surface region decreases with each measurement. Detailed calculations of the flux profiles under oscillatory field perturbations will be presented elsewhere[20]. The size of the flux avalanche decreases for successive measurements as the change in $J_c$ due to the field cycling reduces. Size of the avalanche should also decrease with decreasing $\delta H_a$, which is evident from the data presented in

Fig. 6. During the waiting period, relaxation in the flux gradients due to thermally activated flux creep cannot be ruled out. But it is perceptible only above $H_p$ from $ln(t_0)$ being significantly smaller than unity.

An important conclusion is that the presence of a supercooled metastable high $J_c$ phase is crucial in explaining the athermal relaxation process observed on the reverse magnetization curve. It is widely believed that vortex matter exhibits metastability only below $H_p$. For instance, minor magnetization curves do not reveal any signature of metastability above $H_p$[11,12]. Athermal component of the relaxation ($ln(t_0) \neq 0$ ) observed at fields well above $H_p$ signifies the presence of vortex state metastability. We propose that, analysis of the relaxation data with different time intervals as described above provides a sensitive probe of the metastable vortex states. The discussion presented above is specific to the QD SQUID magnetometer. However, it can be extended to other techniques where sample movement in an inhomogeneous field is employed, with appropriate modifications.

At a given field on the reverse curve, magnetic moment falls by about 30 percent of the initial value after three measurements. Subsequently, magnetic moment is recorded by decreasing the magnetic field. The resulting magnetization curve readily merges with the reverse branch as shown in Fig. 8, irrespective of whether the experiment is carried out below or above $H_p$. This is not expected if the initial and the relaxed states (after three measurements) correspond to different metastable vortex configurations. The avalanche produced by the oscillatory field perturbation does not seem to alter the vortex lattice correlations in the interior of the sample.

Further, relaxation data at fields below $H_p$ (on the reverse curve) have an inflection point as shown in fig. 9. This is reminiscent of time relaxation experiments by Kunchur et al[21]. We speculate that the inflection point occurs at a point when $J_c$ in the surface region attains a stable value.

In the forward case we know that the vortex state corresponds to a superheated state[12,17]. A field perturbation in this case results in an increase in $J_c$ near the surface. Therefore the flux profile in the interior remains unaffected by the perturbation as shown in Fig. 7(b). The relaxation measured in the forward case seems to be due to the decay of induced current in the interior region due to the phenomenon of flux creep.

In conclusion, we measured magnetization relaxation in the peak effect regime of superconducting $V_3Si$ using Quantum Design SQUID magnetometer. On the forward magnetization curve, relaxation behavior is due to thermally activated flux creep. Athermal relaxation observed on the reverse magnetization curve is predominantly governed by the flux avalanches triggered by the field excursion experienced by the sample during a measurement scan. Presence of supercooled disordered phase is essential for the athermal relaxation process.

**Acknowledgements**:

G. Ravikumar thanks Prof. Shobo Bhattacharya, Dr. K. V. Bhagwat, Dr. P. Chaddah and Dr. Mahesh Chandran for discussions.

**Figure captions**:

Fig. 1: Magnetization hysteresis loops in the peak effect regime using 1.6 and 2.0cm scan lengths.

Fig. 2: *m vs t* data at 2.7T on the forward curve measured at intervals 100 *s* and 300 *s*. Inset shows the normalized logarithmic relaxation rate $S = (1/m_0)(dm/dlnt)$ as a function of field.

Fig. 3(a): *m vs* the serial number n measured at 600 and 200 s intervals, at 2.7T on the reverse curve. Inset shows the two data sets plotted as *ln(m)* vs *ln(t)*.

Fig. 3(b): *ln(m) vs ln(t)* data shown in the inset of fig. 3a replotted with the 200 *s* interval data moved to the right by $ln(t_0)$ = 0.99 to make the two data sets collapse onto each other.

Fig. 4: *ln(m) vs ln(t)* at 3.3T measured at intervals 200 *s* and 600 *s*, 200 *s* interval data is moved to the right by $ln(t_0)$ = 0.62 to make the two data sets collapse on to each other. Note only long time relaxation data collapses by this operation.

Fig. 5: $ln(t_0)$ *vs* field *H* on the reverse curve. Solid curve is a guide to eye. Position of the peak in Jc marked by the arrow

Fig. 6: *m vs n* data at 2.4T on the reverse curve measured using 1.6 and 2.0 cm scan lengths respectively at regular intervals of 200*s*.

Fig. 7(a): Flux profile A (dark line with boundary condition $H(x = 0) = H_a - \delta H_a$ ) denotes the profile before a measurement commences. It is modified upto a depth p as the sample is moved from $z = -l$ to $z = 0$ (see the dark line with boundary condition $H(x = 0) = H_a$). B denotes the profile (dotted line) at the end of one measurement scan when the sample undergoes one complete field cycled of amplitude $\delta H_a$.

Fig. 7(b): Evolution of flux profile during the measurement scan in the forward case.

Fig. 8: *m – H* curves measured by decreasing the field after three measurements at –T and –T on the reverse curve. *m-H* curves readily merge with the reverse curve both below and above $H_p$.

Fig. 9: *ln(m)* vs *ln(t)* data measured at 200 *s* and 600 *s* time intervals with the 200 *s* data shifted to the right by $ln(t_0)$ = 1.02. Note the inflection point indicated by the arrow.

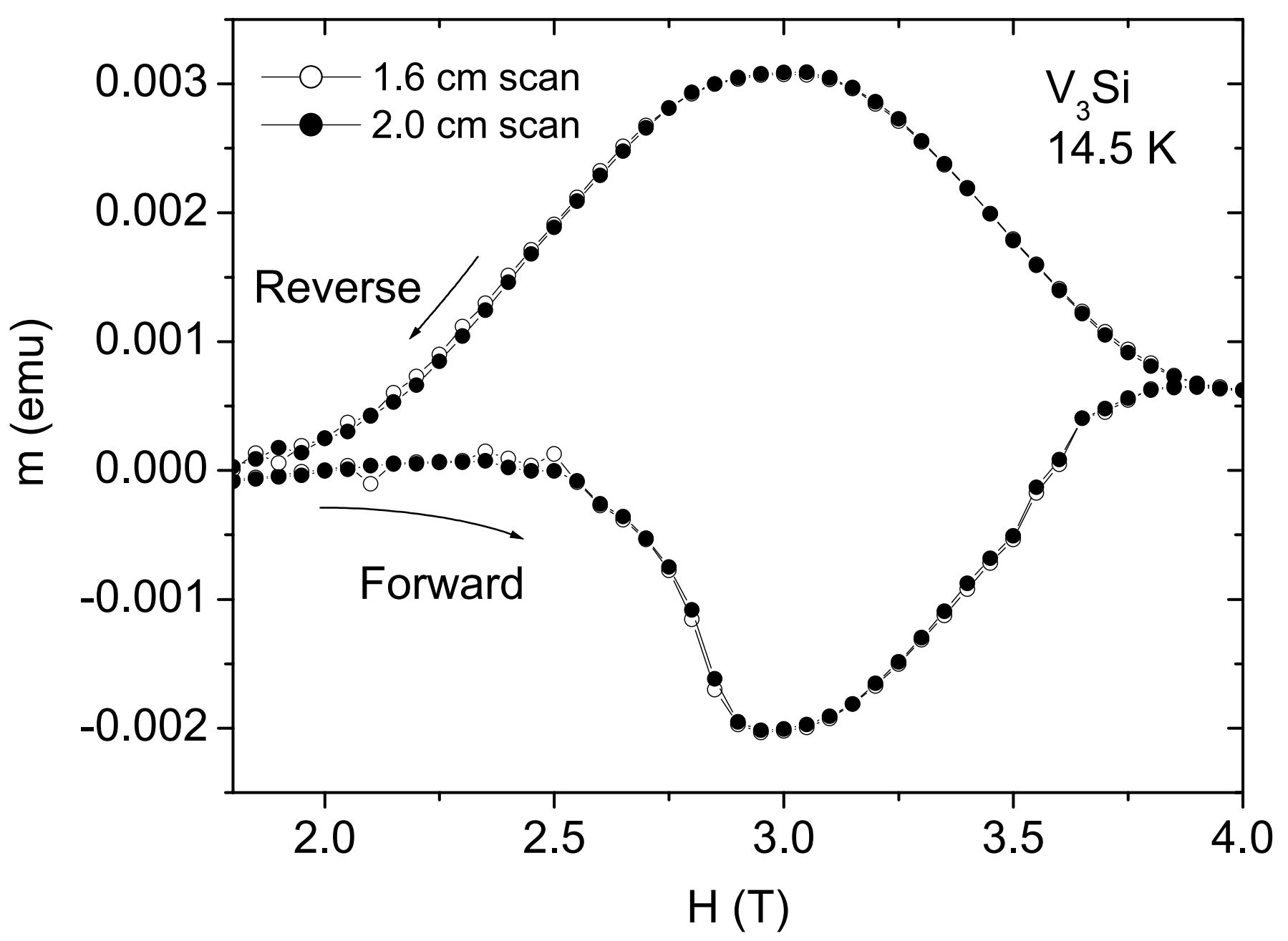

Fig. 1 - Ravikumar et al

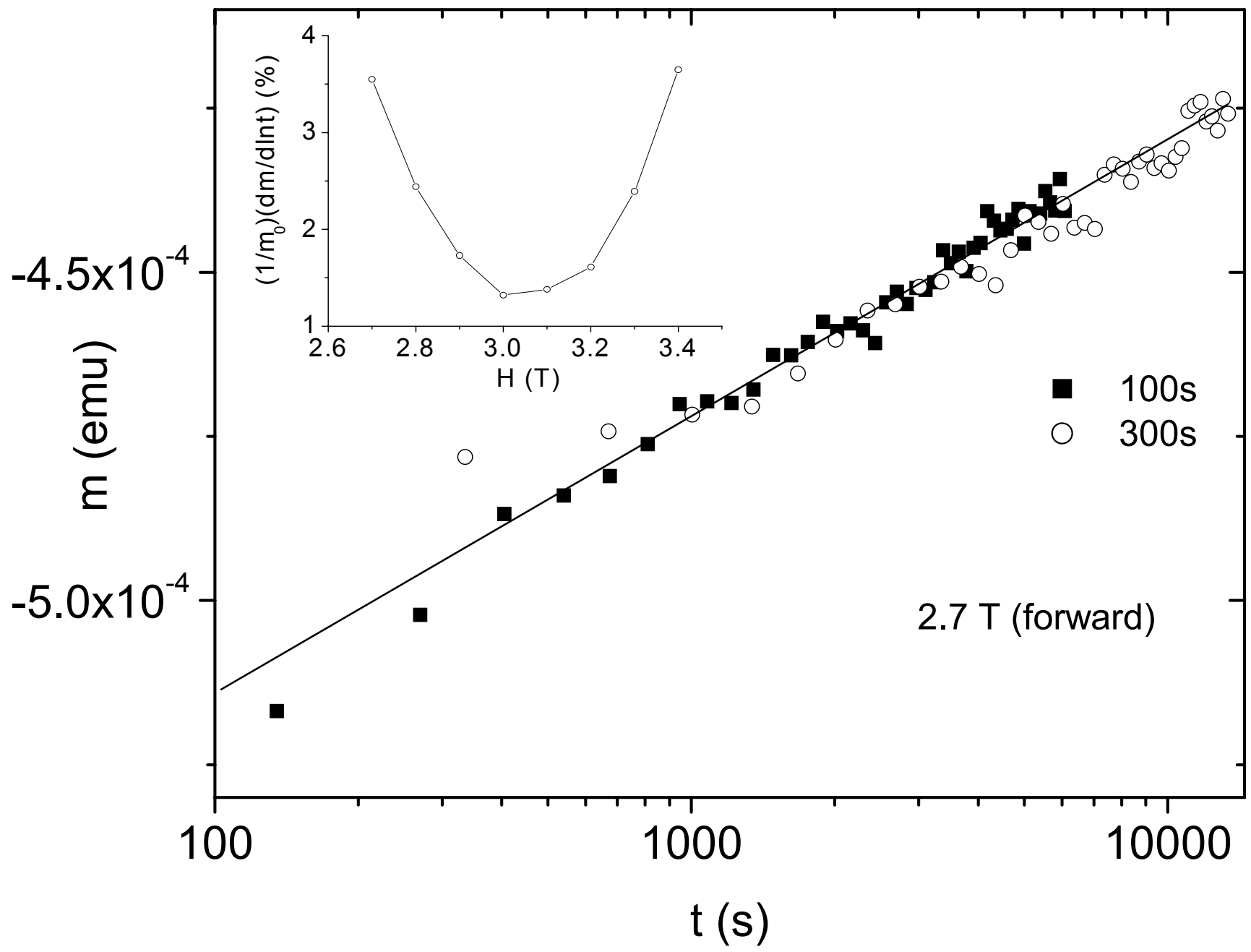

Fig. 2 - Ravikumar et al

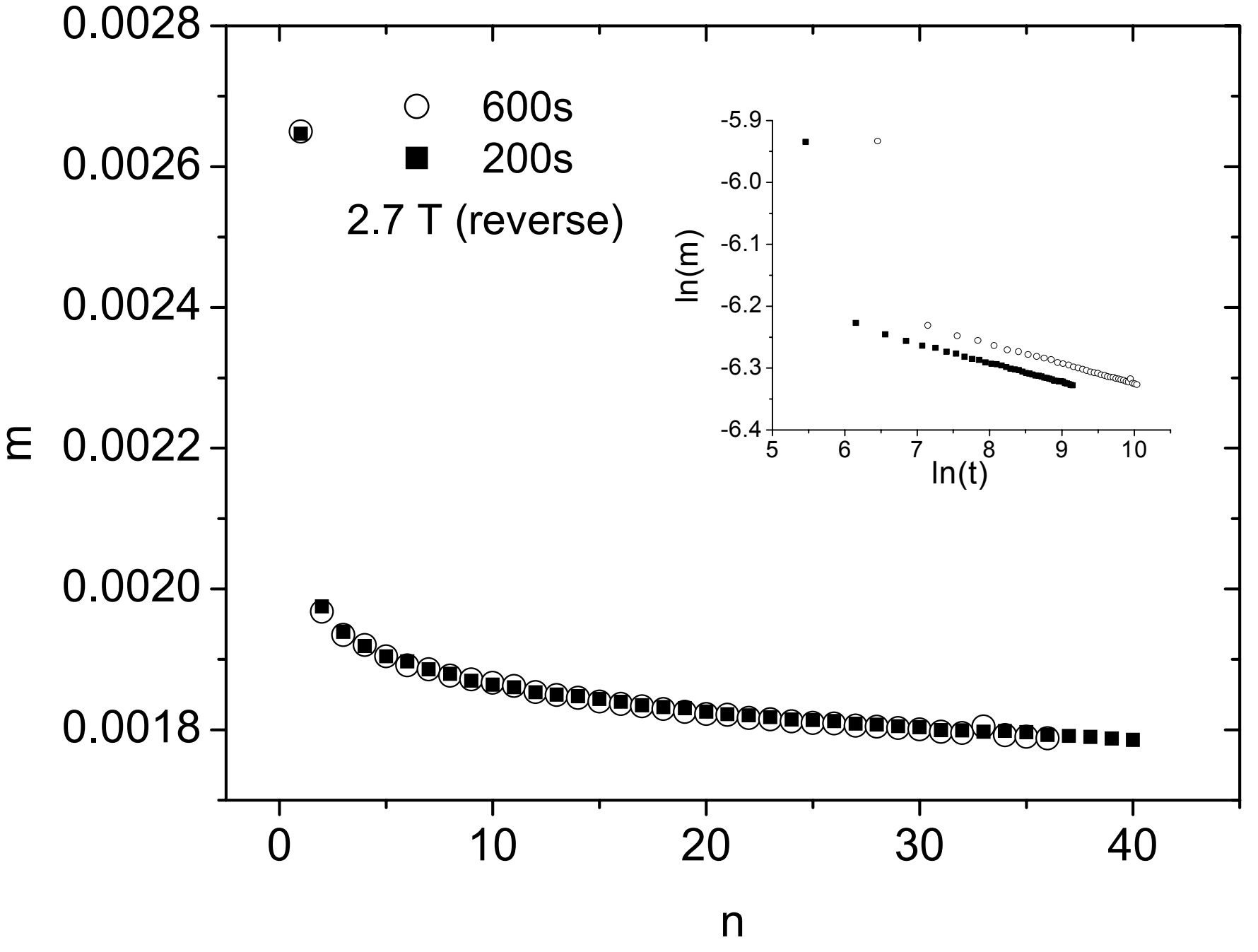

Fig.3a - Ravikumar

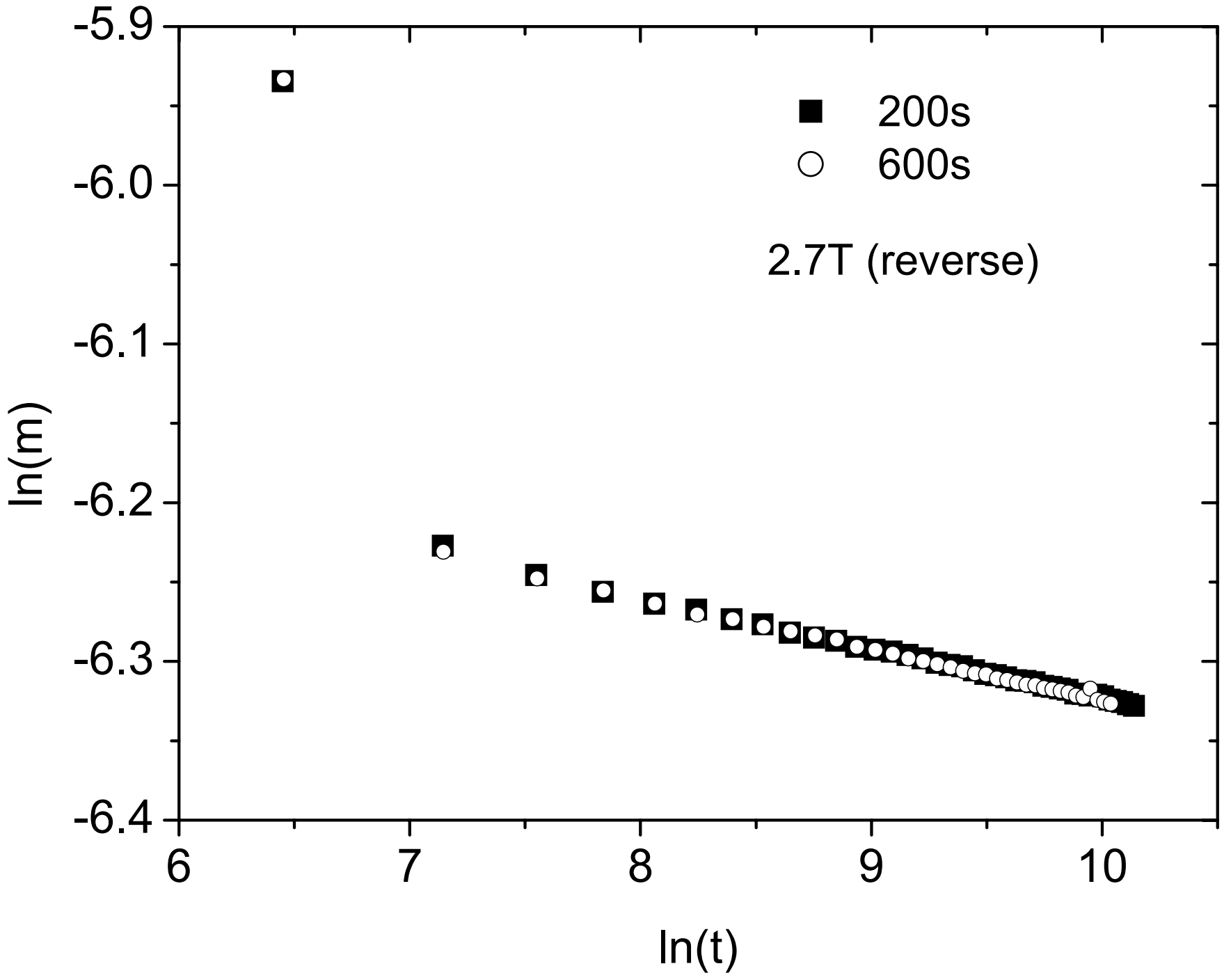

Fig.3b - Ravikumar et al

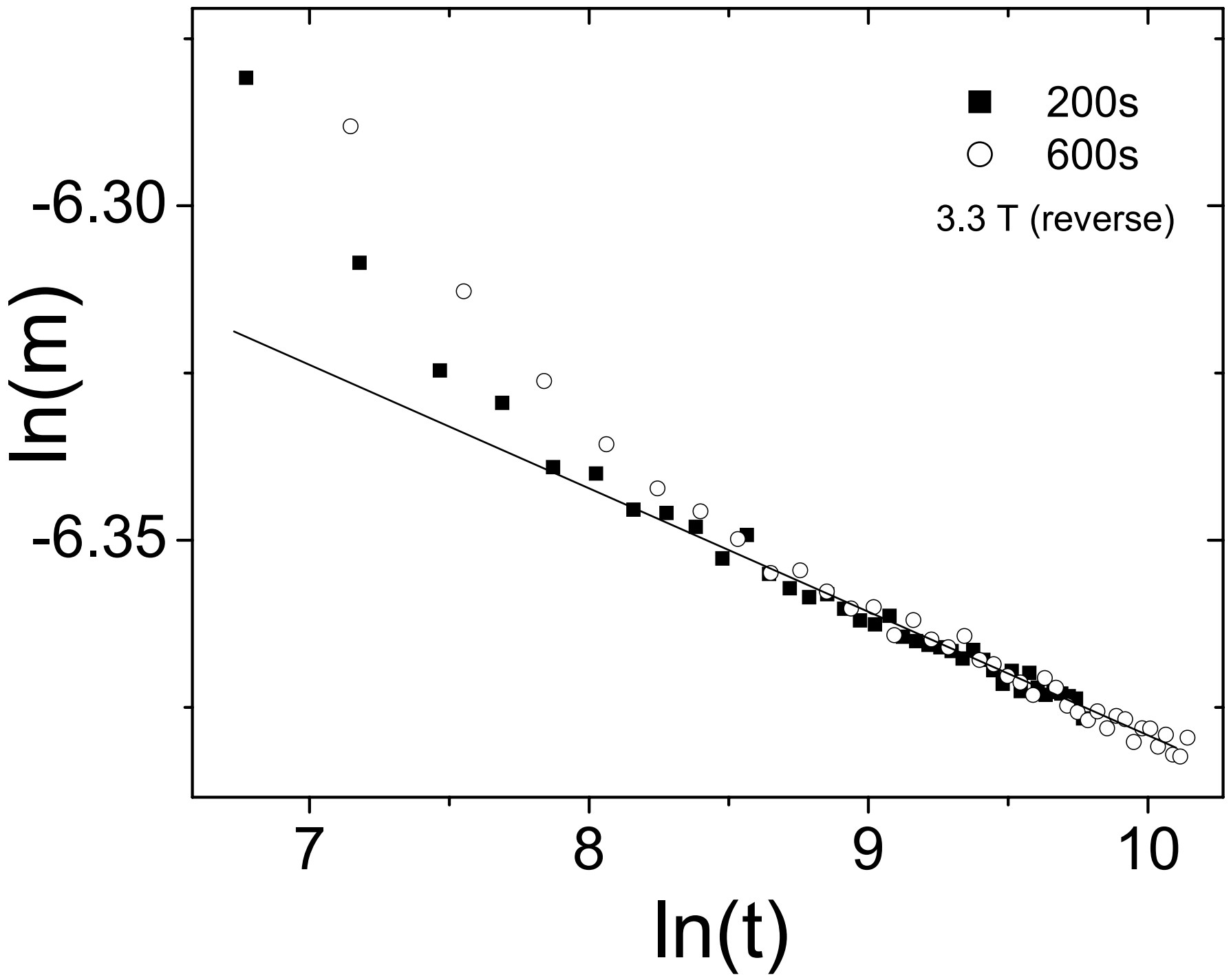

Fig.4 - Ravikumar

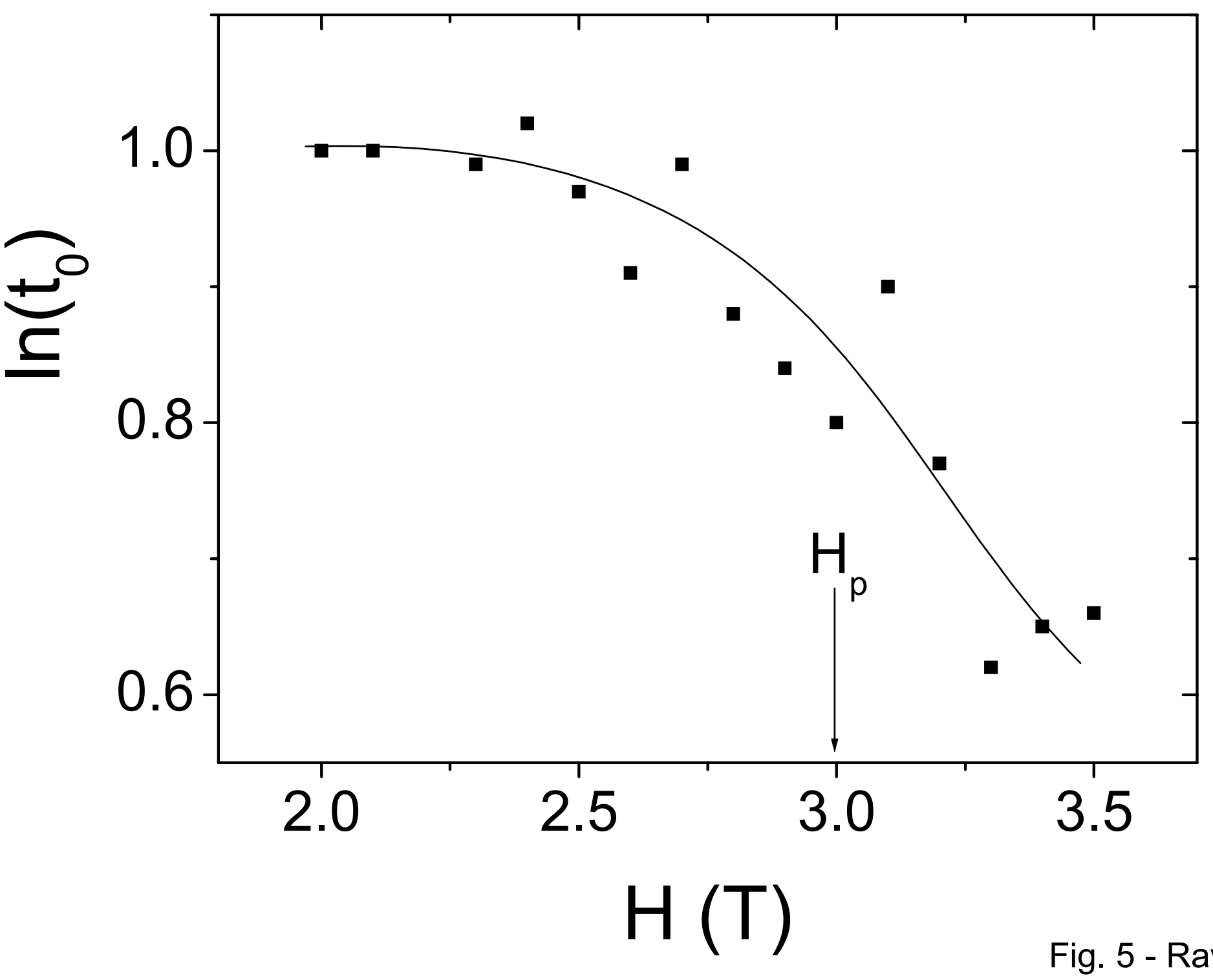

Fig. 5 - Ravikumar et al

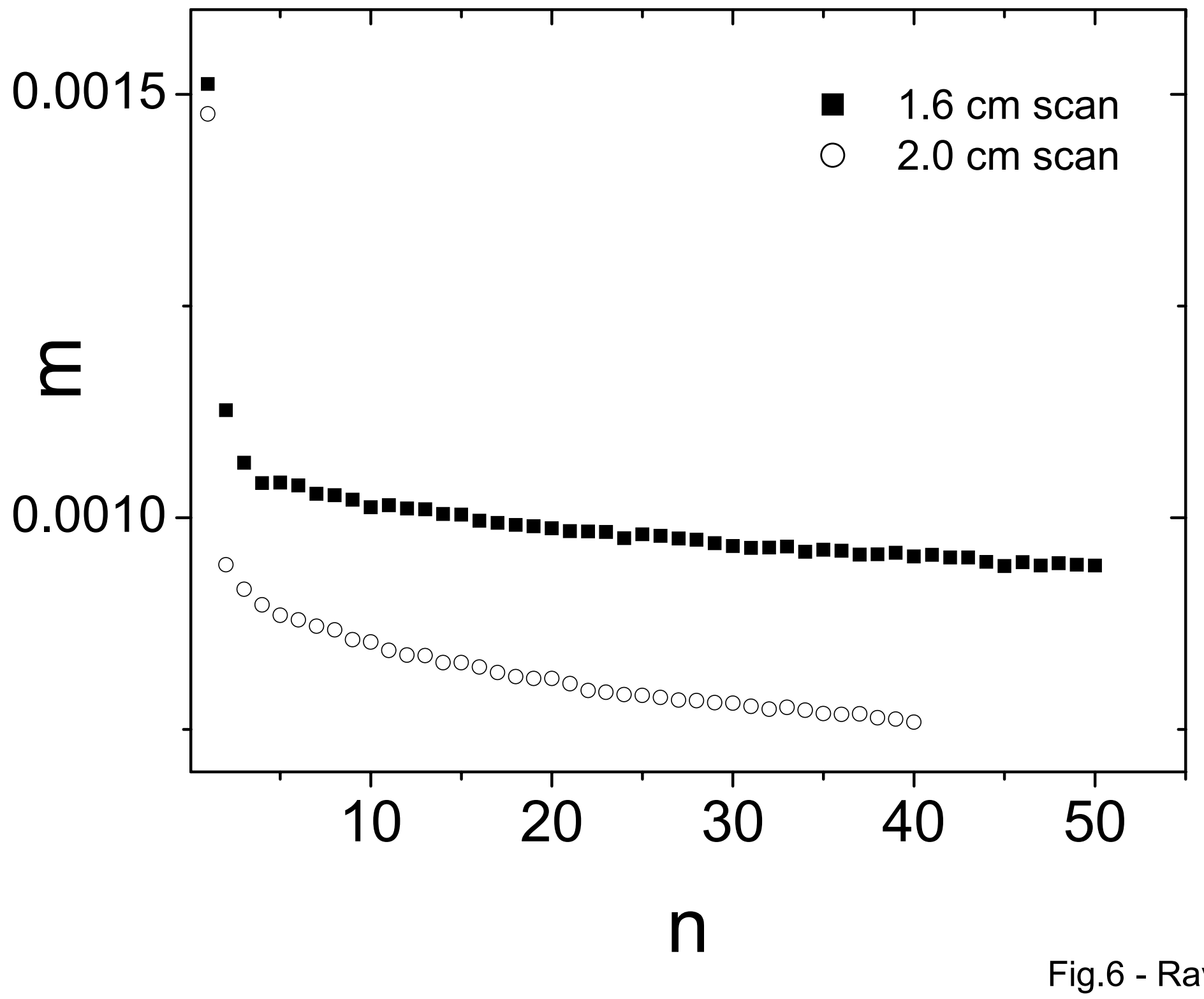

Fig.6 - Ravikumar et al

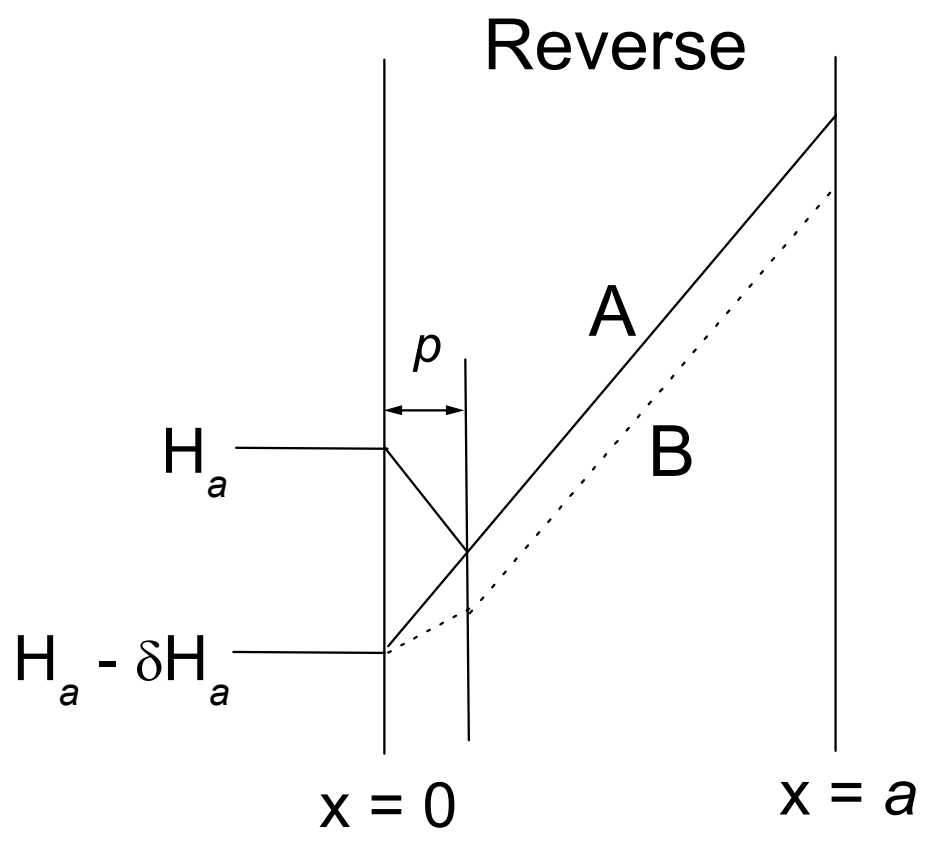

fig.7a - Ravikumar et al

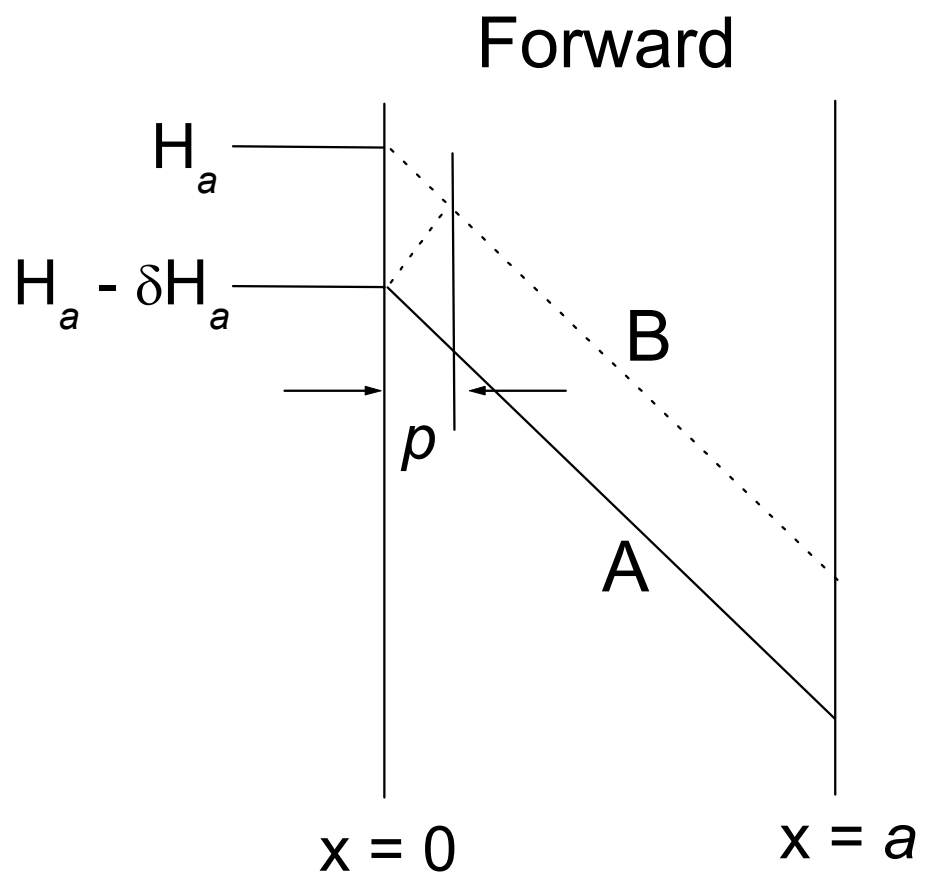

fig.7b - Ravikumar et al

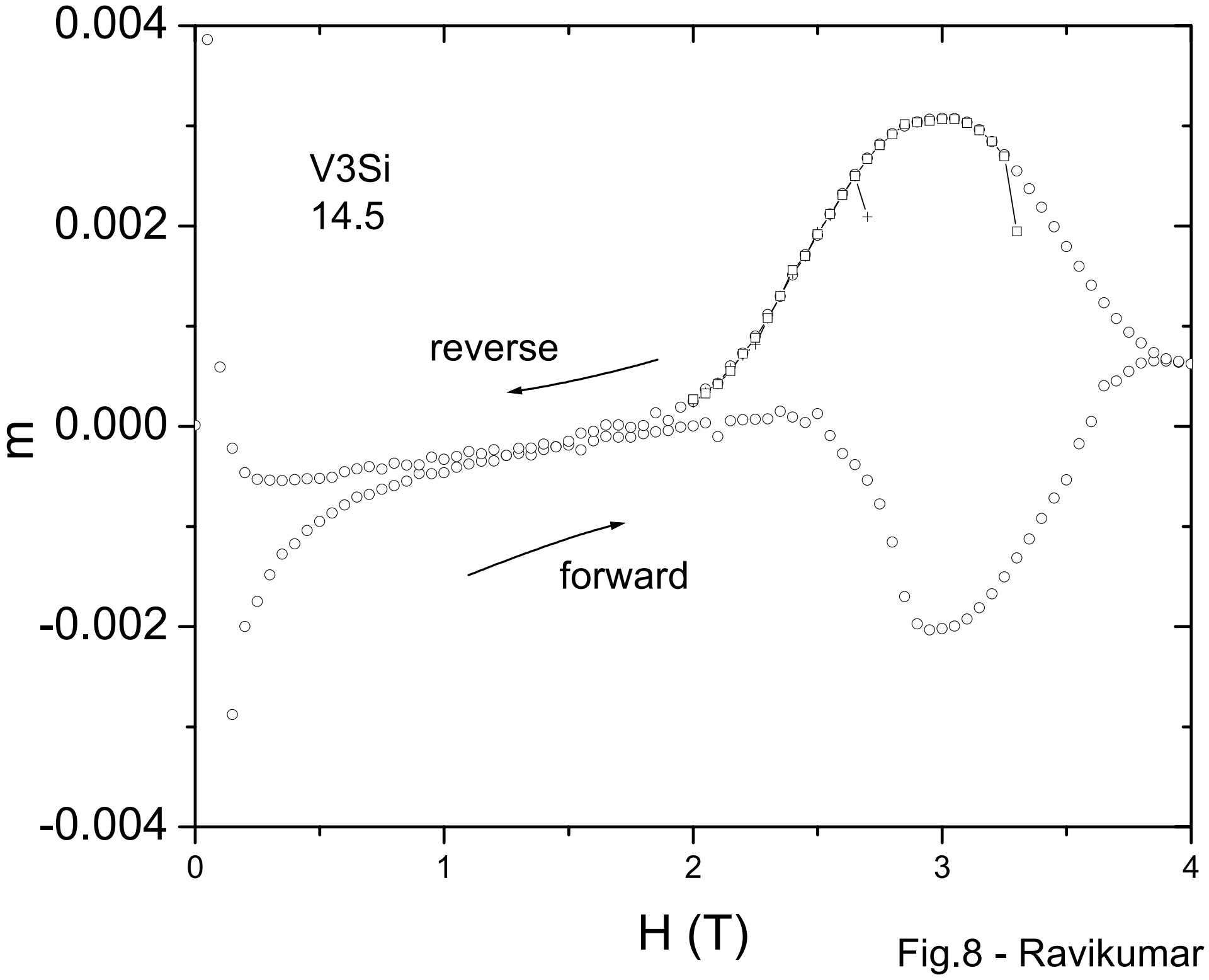

Fig.8 - Ravikumar et al

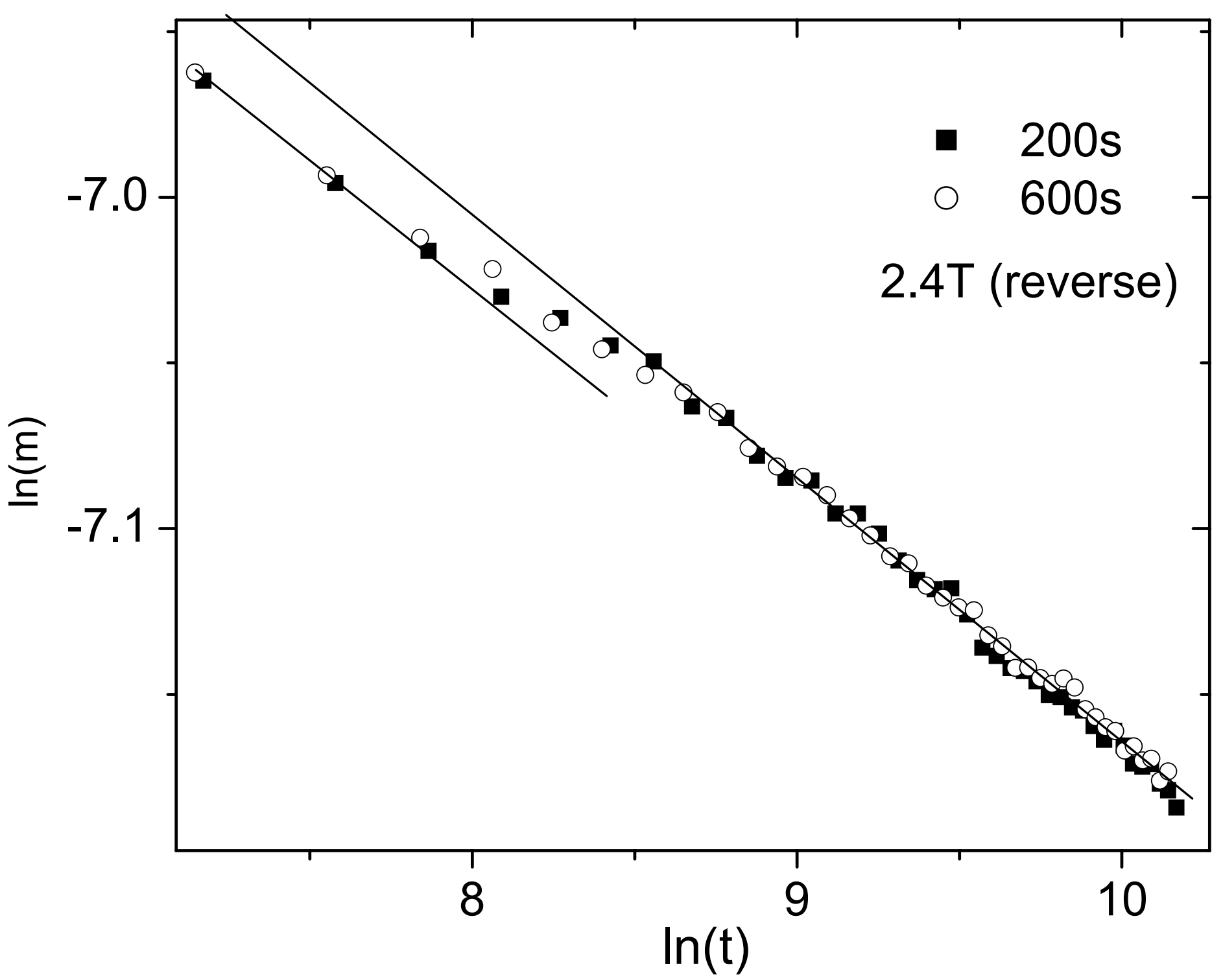

Fig.9 - Ravikumar et al